\documentclass{aa}
\usepackage{epsfig}
\usepackage{graphics}

\newcommand{\be}{\begin{equation}}
\newcommand{\ee}{\end{equation}}

\begin{document}

\title{The Hubble diagram for a system within dark energy:  influence of some relevant quantities}

\author{J. Saarinen\inst{1} \and P. Teerikorpi\inst{1}}

\institute{Tuorla Observatory, Department of Physics and Astronomy, University of Turku, FI-21500 Piikki\"o,
Finland}

\authorrunning{Saarinen and Teerikorpi}
\titlerunning{The Hubble diagram for a system within dark energy:   influence of relevant parameters}

\date{Received / Accepted}

\abstract
{}
{
We study the influence of relevant quantities, including the density of dark energy (DE), to the predicted Hubble outflow
around a system of galaxies.  In particular, we are interested in
the difference between two models: 1) The standard $\Lambda$CDM model, with the everywhere constant DE density, and 2) the "Swiss cheese model", where the universe is as old as the standard model, but the
DE density is zero on short scales, including the environment of the system.
}
{
We calculate the current predicted outflow patterns of dwarf galaxies around the Local Group-like system, using different
values for the mass of the group, the local dark energy density, and the time of ejection of the dwarf
galaxies, treated as test particles. 
These results are compared with the observed Hubble flow around the Local Group.
}
{The predicted distance-velocity relations around galaxy groups
are not alone very sensitive indicators of the dark energy density, due to the obsevational scatter and the uncertainties caused by 
the used mass of the group and  a range in the ejection times. In general, the Local Group outflow data are in agreement
with the local DE density being equal to the global one, if the mass is  about $4 \times 10^{12} M_{\odot}$; a lower mass $\la 2 \times 10^{12} M_{\odot}$ could suggest a zero local DE density.
The dependence of the
inferred DE density on the mass is a handicap in this and other
common dynamical methods. This emphasizes the need to 
use different approaches together, for constraining
the local dark energy density.   
}
{}


\maketitle

\section{Introduction}

 Peirani \& de Freitas Pacheco (2008) and Chernin et al. (2009) made calculations on the present-epoch Hubble relation around galaxy groups, without and with local dark energy (DE) and pointed out differences expected in these two cases. For instance, with dark energy,
the zero-velocity distance $R_0$ will be longer than without dark energy, for a fixed group mass.
Teerikorpi \& Chernin (2010) further considered, within the Einstein-Straus vacuole containing different amounts of dark energy, the expected present-day local Hubble relation and its link to the global Hubble law. 

Here we extend the previous works to include a range of relevant parameters (mass, ejection time) in order to reach better understanding on
the factors influencing the use of the local Hubble flow as a measure of the dark energy. 
 We concentrate in the cases where the central mass (point mass or binary), keeps constant, in order to see clearly the basic predictions and problems which would also go over into any perhaps more realistic and still uncertain models (such as having a changing mass).

Along with the standard $\Lambda$CDM model, with its constant dark energy density on all
scales, it is relevant consider the "Swiss cheese model", where the universe has the same age as the standard model, but the DE density is zero on short scales.
 This could correspond to the case where dark energy (or analogous effects) operates on large scales only. 

\section{Dark energy on local scales}

Our starting point is the outflow model (Chernin 2001; Chernin et al. 2006; Byrd et al. 2012)
intended to describe the major features of
expansion flows around local masses, which was motivated by
the observed picture of the Local Group
with outflowing dwarf galaxies around it
(e.g., van den Bergh 1999; Karachentsev et al. 2009).
 The model treats the dwarfs as "test
particles" moving in the force field produced by the
gravitating mass of the group and the possible dark energy
background.

In order to see the main expected features, one may 
consider a spherical "vacuole", as
given by the exact Einstein-Straus solution (Einstein \& Straus 1945; Chernin et al. 2006). The vacuole
may be seen as defining the region from
which the central mass $M$ (the galaxy group) has been gathered. In the vacuole, 
the K$\ddot o$ttler solution gives an exact
description of a spherically-symmetrical spacetime (the Schwarzschild-de Sitter spacetime) outside a spherical
mass $M$ within dark energy of constant density
$\rho_{\rm de}$.   
When the gravity/antigravity field is weak,
one may write (Chernin et al. 2006)
%
the force (per unit mass) as
the sum of the Newtonian gravity force
produced by the mass $M$ and the
Einstein force of antigravity due to dark energy:
\be F(R) = - \frac{GM}{R^2} + \frac{8\pi G}{3}
\rho_{\rm de} R. \ee

A gravitationally bound group is located within its zero-gravity
sphere having the radius $R = R_{\rm ZG}$: 
\be R_{\rm ZG} = (\frac{M}{\frac{8\pi}{3} \rho_{\rm de}})^{1/3}, \ee
\noindent (Chernin 2001).
At this
distance, gravity $=$ antigravity.

If the mass is, say, $ 2 \times 10^{12} M_{\odot}$
and the local dark energy density is equal to its global value $\rho_{\rm v}$,
$\rho_{\rm v}  \approx 7 \times 10^{-30}$ g/cm, 
 then
$R_{\rm ZG} = 1.3$ Mpc.

\section{Local outflow}

The radial motion of the flow is
controlled by the force field given by Eq.1, and the
equation of motion at $R \ge R_{\rm ZG}$ is
\be \ddot R(t) 
= - \frac{GM}{R^2} + \frac{8\pi G}{3}\rho_{\rm de} R. \ee
\noindent
The first integral of this equation is
\be \frac{1}{2} \dot R^2 = \frac{GM}{R} + \frac{4\pi G}{3}
\rho_{\rm de} R^2 +E. \ee
\noindent $E$ is the constant total mechanical energy of the particle.

%
%

For a particle to escape from the potential well of the group (Chernin et al.
2006), its energy $E$ must exceed
$E_{\rm esc} = - \frac{3GM}{2R_{\rm ZG}}$,
which defines the lowest possible velocity in the outflow.
Because the ejection happened a finite time ago, the lowest velocities are generally  higher than this.
%
%

The local flow is generally not linear which is due to the non-uniform,
point-like matter distribution, and the now observed Hubble flow
consists of objects having different energies $E$,
while in the global flow of uniform matter, the trajectories are parabolic (the total
energy $E = 0$ in Eq.4).

If, say, two objects with equal energies were expelled at different
past times, they should still now lie on the trajectory corresponding to
that same energy. However, if the objects had a range of different energies when expelled at the
same past time, their now-observed distance-velocity locus
is made of pieces of trajectories for different $E = E(R)$
(Chernin et al. 2009; Teerikorpi \& Chernin 2010). The true situation could be a mixture of these variants:
a range of energies, a range of expulsion times.

The present-day outflow of the objects if expelled at the same time
thus  depends on the mass of the group, the flight time  ($\la$ the age of the universe), and
the local DE density.

\section{The method of calculation}

Peirani \& de Freitas Pacheco (2008) derived the velocity-distance relation using numerically the Lema$\hat{i}$tre-Tolman model containing the cosmological constant, and compared this with the LT-model result for $\Lambda = 0$, using the central
mass and the Hubble constant as free parameters.
Chernin et al. (2009) constructed the expected present Hubble relation by considering
the total energy for different outflow velocities at a fixed distance from the group, and then
calculating the required time for the test particle to fly from near the group's center of mass up to
this distance. The locus of the points corresponding to a constant age ($\approx$ the age of
the universe) gave the expected relation.

Here we use a variant of the above methods which easily allows one to vary the values of the relevant parameters and to generalize
the model in various ways. We generate particles close to the center of mass of the group and
give them a distribution of speeds. Then we let them fly along the radial direction for a time $T$,
and their distances from the center are noted. The flight time $T$, during which the integration of the equations of
motion is performed. is at most the age of the standard
universe, 13.7 Gy. The locus of the points corresponds to the distance-velocity relation as observed
at the present time.

Figure 1 illustrates the method by showing the relevant kinematial curves for two background models with the same age. We see that
in the standard model the particles reach larger distances, while in the Swiss Cheese model the distances are shorter because
only gravitation is acting. We also see that individual particles can reach quite different distances even for the same initial conditions.
The final present-day relations also differ, but not very much. The difference is in the sense already noted by Peirani \& de
Freitas Pacheco (2008) and Chernin et al. (2009): models with no dark energy tend to give a shallower local Hubble relation and
a longer zero-velocity distance.

\begin{figure}
\epsfig{file=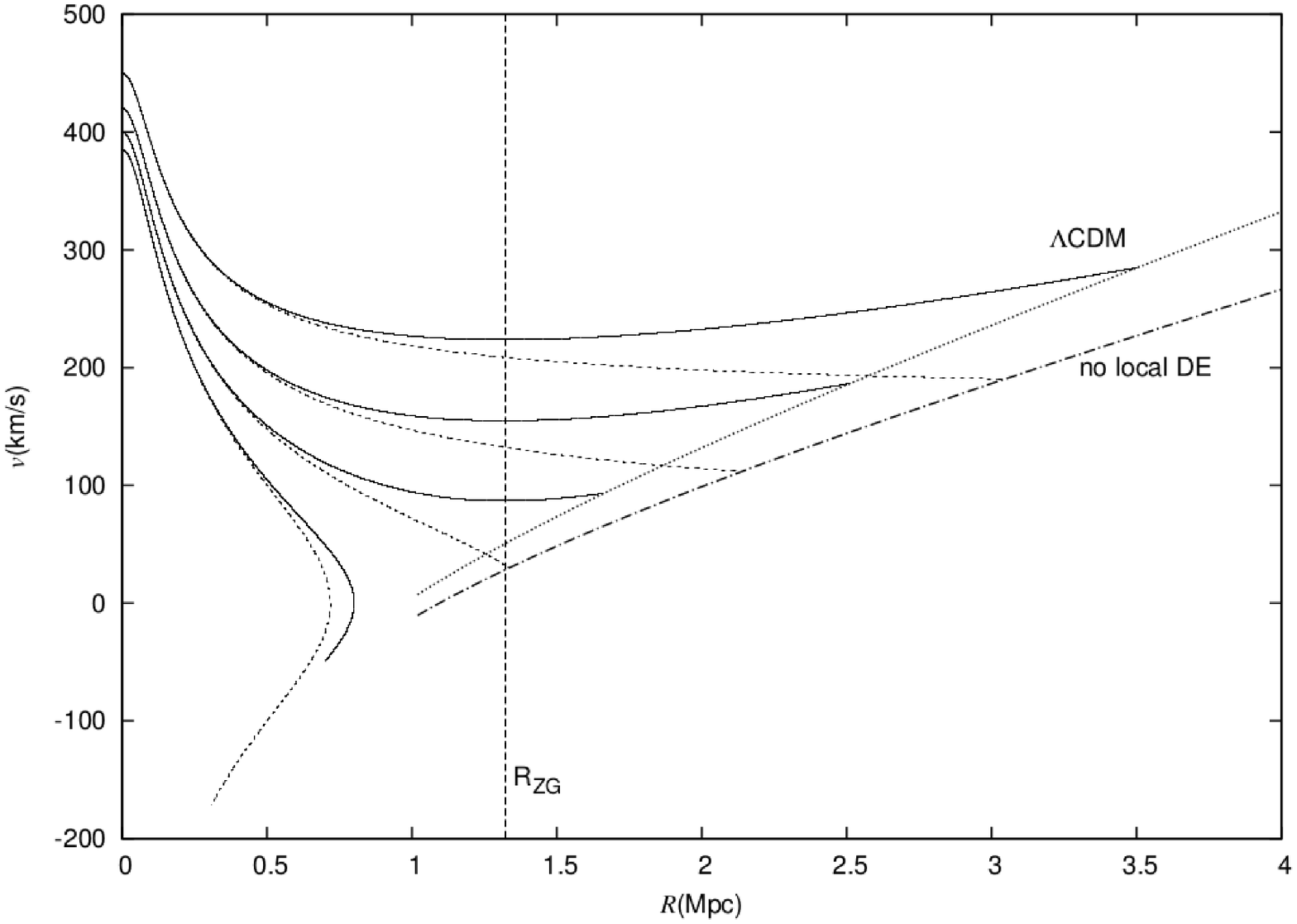, angle=0, width=8.0cm}
\caption{Trajectories of test particles as injected from the mass centre (continuous curves). The particles are initially given different velocities and then they are allowed to travel for a given time $T_0$ (here $T_0 = 13.7$ Ga). The dashed curves show where they
are located at the time $T_0$. Red curves refer to the standard model (with dark energy on all scales), the green ones indicate
the results for the Swiss Cheese model (no dark energy on local scales). The vertical dashed line is the zero-gravity distance, Here
the central mass is $ 2 \times 10^{12} M_{\odot}$. Note the minimum velocity at $R = R_{\mathrm ZG}$ for each $\Lambda$CDM
model trajectory. 
}
\label{fig1}
\end{figure}

\begin{figure*}
\centering
\epsfig{file=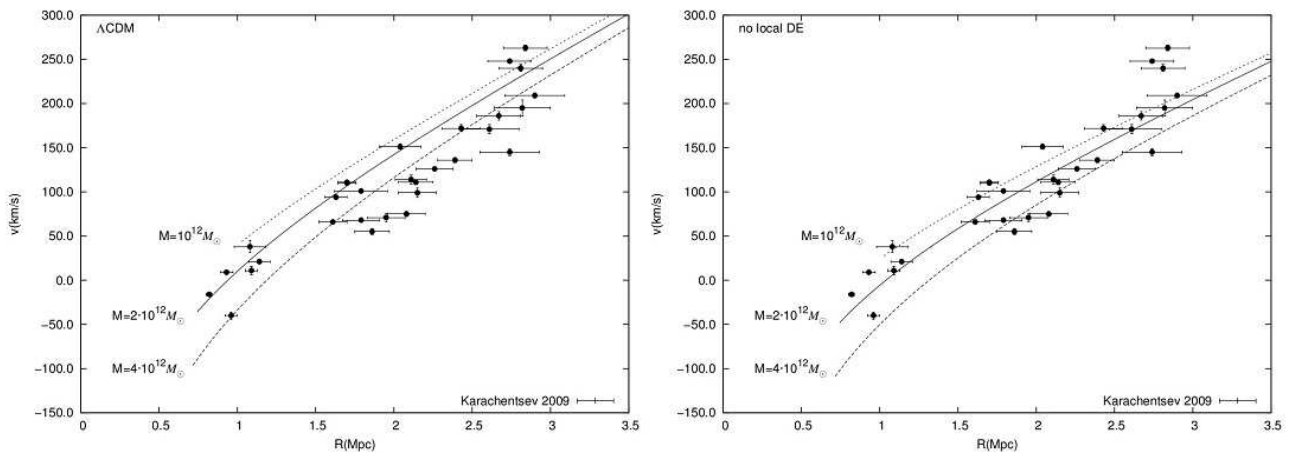, angle=0, width=17cm}
\caption{Left  panel: The location of test particles as injected from the mass centre (curves) after the flight time $T_0 = 12.7$ Ga,
for different masses ($ 1 \times 10^{12} M_{\odot}$,$ 2 \times 10^{12} M_{\odot}$,$ 4 \times 10^{12} M_{\odot}$).
Here the standard model is used (local DE = global DE).
Right panel: The same for  the Swiss cheese model (no local dark energy).
The data points are for the Local Group. 
}
\label{fig_multimass_de_rj}
\end{figure*}

\section{Different models and parameter values: results}

The case where the particles have left the center just after the Big Bang would correspond to a classical situation where the outflow around the central group is "primordial". We show also some results for 1 and 2 Gy after the Big Bang. 
In practice, the age of the group and the outflowing dwarf galaxies must be less than the age of the Universe, and the origin of the outflow may be due to early interactions within the system, making galaxies escape from it (e.g., Byrd et al. 1994, 2012; Chernin et al. 2004). The dark energy antigravity enhances the escape probability because it makes the
particle potential energy barrier lower than in the presence of gravity only.



We plot the results together with the data on the local outflow around the Local Group
from the Hubble diagram as 
derived  by Karachentsev et al. (2009) from   HST observations. 
The largest available distance is 3 Mpc.

\subsection{Point-mass model: influence of mass}

Figure 2 shows the predicted distance--velocity curves for different masses in the $\Lambda$CDM model and
for the Swiss Cheese model. 
The location of the curve is seen to depend rather strongly on the adopted mass. A higher mass makes
the location of the curve lower (or increases the zero-velocity distance). 
A factor of two uncertainty in the used mass makes it difficult to decide between the standard model and
the locally zero DE model (SwCh).
For instance, using a too low value of mass, can lead one to conclude that there is no local dark energy, even if
there is.

Of the models shown, only the SwCh model with $ M = 4 \times 10^{12} M_{\odot}$
and the standard model with $ M = 1 \times 10^{12} M_{\odot}$ seem to be excluded in light of
the Local Group data.
By eye, the SwCh curve with $ M = 2 \times 10^{12} M_{\odot}$  gives a good fit for the LG data along
the whole distance range. In fact, the most likely fitting mass for the SwCh model is
$ 1.5\times 10^{12} M_{\odot}$, as determined by the least squares method.
 The $\Lambda$CDM model requires $ M \approx 4 \times 10^{12} M_{\odot}$ for a good
fit beyond the zero-gravity distance (as was already noted by Chernin et al. (2008). 
The results illustrate the importance of knowing in some independent way the true mass.



\subsection{Point-mass model: influence of ejection time}

Figure 3 shows the predictions for the Hubble relation when the test particles have been ejected at three different
times, 0, 1, and 2 Ga after the Big Bang. Later ejected particles will have higher velocities, if observed now. This corresponds to the inverse relation between the global Hubble constant and the Hubble time in Friedmann models.  

The later the particles have been ejected, the lower should 
the local dark energy density be for the prediction to agree with the data for a fixed mass.

\subsection{Binary model}

Actually,  the Local Group has two main components, the Milky Way and M31.
It is in fact this binary structure which has been suggested as a possible source for the ejection of dwarf galaixes (Chernin
et al. 2004). On the other hand, the zero-gravity surface as calculated for the Local Group is almost spherically
symmetric at relevant distances (see Fig.1 in Chernin et al. 2009), so one would not expect a large effect.

We have checked the influence of the binary structure on the distance--velocity relation by comparing the point-mass
result with the binary structure result in two extreme directions: perpendicular to the components and parallel to the
components. Appendix gives the used force laws.

Figure 4 shows that the expected spread due to the binary mass point structure is indeed rather small in comparison with the other
factors.

\begin{figure}
\epsfig{file=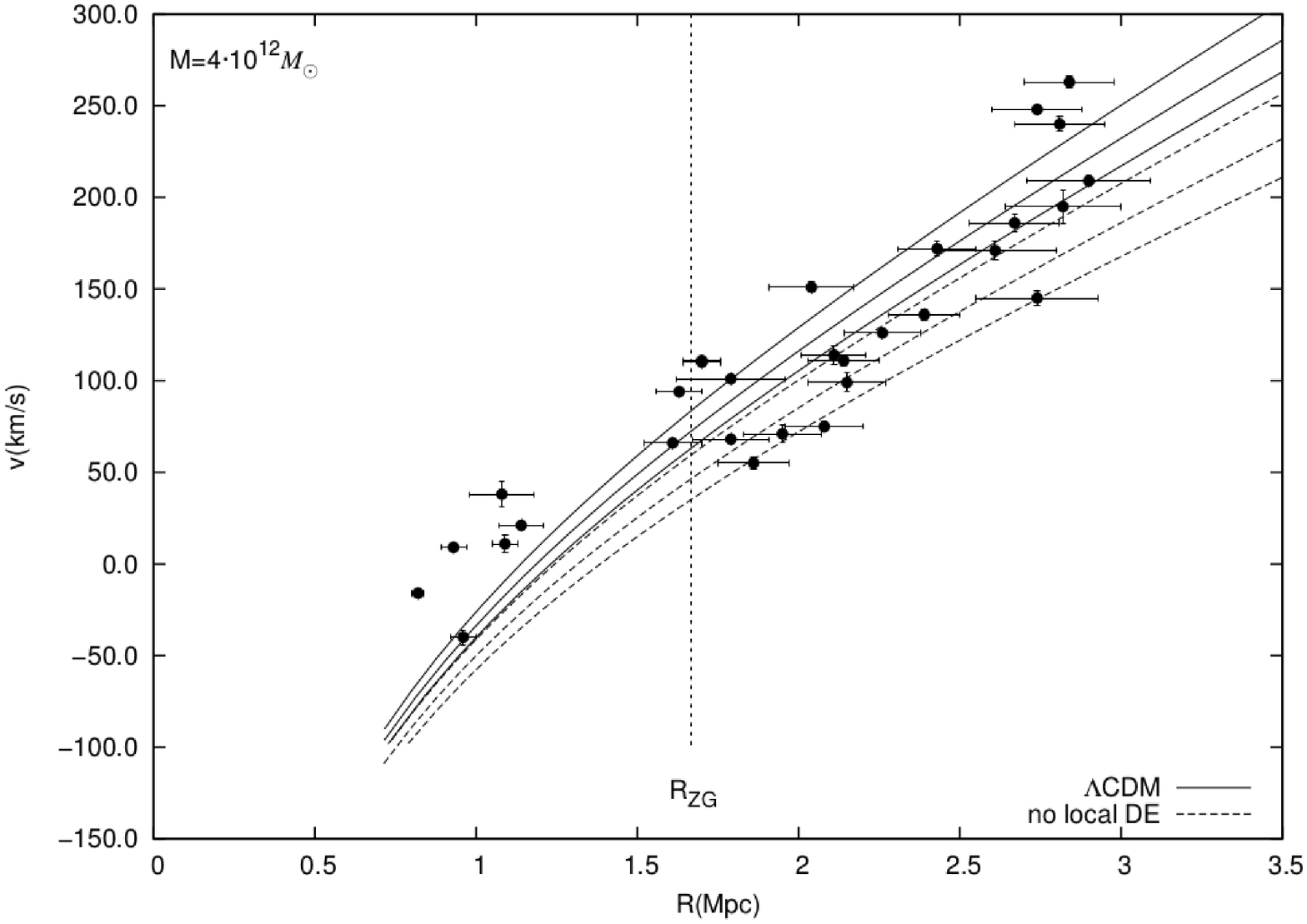, angle=0, width=8.0cm}
\caption{The location of test particles as injected from the mass centre (curves) after the flight times $T_0 = 13.7$ Ga,
$T_0 = 12.7$ Ga, and $T_0 = 11.7$ Ga
for the  mass $ M = 4 \times 10^{12} M_{\odot}$.
A spread in ejection times will increase the scatter. Particles which have been ejected later are found on
higher trajectories. The lowest curves of the two triplets correspond to the age of the Universe.
}
\label{fig4}
\end{figure}

\begin{figure}
\epsfig{file=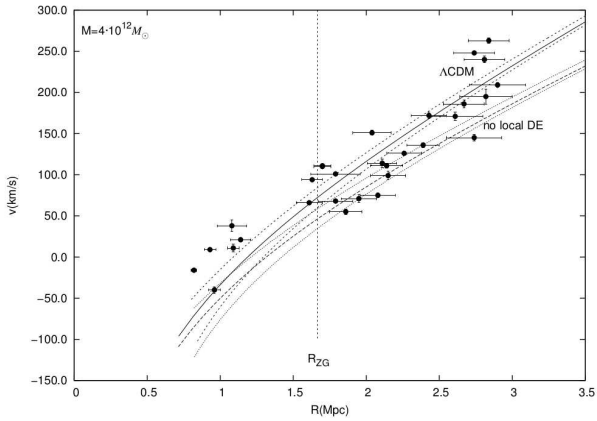, angle=0, width=8.0cm}
\caption{The location of test particles as injected from the mass centre after the flight time
$T_0 = 12.7$ Ga for the point  mass $ M = 4 \times 10^{12} M_{\odot}$ (the solid curves) and for the binary structure
in perpendicular direction (the upper dashed curve) and in parallel direction (the lower dotted curve).The upper triplet of curves corresdonds to the standard
model, while the lower triplet is for the Swiss Cheese model.
}
\label{fig5}
\end{figure}

\section{Discussion and conclusions}



Our calculations show clear differences between the cases of $\Lambda$CDM and no local dark energy, in the sense expected. In practice, the difference may be difficult
to detect.  First, the observed distance--velocity relation is rather scattered, roughly covering a similar region in the
$r-V$ diagram as the theoretical curves. Second, 
the independently known masses of the galaxy groups (even our Local Group) are uncertain. Thirdly,
the evolution of a galaxy group contains uncertain elements, including the ejection time.

We saw that the Hubble flow around the Local Group to be approximately consistent with the condition $\rho_{\rm loc} = \rho_{\Lambda}$
requires a rather high mass $\approx 4 \times 10^{12} M_{\odot}$. The usually cited $\approx 2 \times 10^{12} M_{\odot}$
would suggest about zero local DE density.
Indeed, one point to emphasize is the interdependence between the assumed mass and the inferred DE density. This occurs here and also in other
common dynamical methods such as the virial theorem and the Kahn-Woltjer method (Chernin et al.  2009).
 
In fact, attempts to probe dark energy with nearby outflows were
first made (e.g., Chernin et al. 2006; Teerikorpi et al. 2008) by using the zero-gravity
radius (Eq.2). This method is also affected by the "mass--dark energy degeneracy". If the possible range of $R_{\rm ZG}$
 is somehow estimated from the velocity-distance diagram\footnote{According to our
calculations the Hubble relation of the outflow does not show any clear signature of the zero-gravity distance. 
However, Teerikorpi \& Chernin 2010 showed that
if the outflow reaches the global Hubble rate at a distance $R_2$
from the group, the zero-gravity radius is expected to lie at $\la R_2/1.5$. For
the Local Group this is roughly where
the outflow begins to be seen.
}
and the mass $M$ of the group is independently known or assumed, the
DE density may be estimated in the outflow region:
\be \frac{\rho_{\rm loc}}{\rho_{\Lambda}} = (\frac{M}{2 \times 10^{12}
M_{\odot}}) (\frac{1.3 {\rm Mpc}}{R_{\rm ZG}})^3.\ee
Here the DE density is directly proportional to the mass.
Hence Eq.5 may give useful constraints on the DE density
for methods where the  $\rho_{\rm loc}$ vs. $M$ dependence is shallower,
such as the  modified Kahn-Woltjer method, which results in a non-horizontal elongated area of admissible values in the local DE density-mass diagram. Thus Chernin et al. (2009) derived using
the relative velocity and distance of our Galaxy and the M31 galaxy together with the outflow data the local
DE density as $\rho_{\mathrm loc}/\rho_{\Lambda} = 0.8-3.8$.
Using a similar approach, Byrd et al. (2014) revised this range to $\rho_{\mathrm loc}/\rho_{\Lambda} \approx 0.5-2.5$.

To conclude,

$\bullet$ The outflows around galaxy groups, though theoretically clearly influenced by local dark energy,
are not very sensitive indicators of the exact value of the dark energy density in view of the obsevational scatter in the distance-velocity diagrams, as already noted by 
Peirani \& de Freitas Pacheco (2008), and
the uncertainties caused by
the adopted mass of the group, a range in the assumed ejection times, and the binary nature of the group. Especially the uncertainties in mass and time are harmful.

$\bullet$   In general, the Local Group outflow data are in agreement
with the local DE density being equal to the global one, if the mass is  about $4 \times 10^{12} M_{\odot}$. A lower mass $\la 2 \times 10^{12} M_{\odot}$ could suggest a zero local DE density.

$\bullet$ Theoretically, local dynamics on galaxy group (a few Mpc) scales significantly differs for the cases of local DE = global
DE or = 0 (e.g., Chernin et al. 2009, Byrd et al. 2012), though any  single method (e.g. local outflows, the Kahn-Woltjer
method, the virial theorem) have large uncertainties. Hence, it is desirable to consider different methods together in order
to lower systematic and random errors.   

 Up to now, no local evidence has appeared contradicting
the concept of the universal cosmological constant as adopted in the
$\Lambda$CDM cosmology. Of course, such evidence if found
would be very challenging for the standard cosmology (see, e.g., Byrd et al. 2012).

{\it \bf Acknowledgements}  We are grateful for discussions with G. Byrd,
A.C. Chernin, and M. Valtonen.

\appendix
\section{Force laws in the binary model}

In the binary model the perpendicular force is taken to have the form

\begin{equation}
\vec{F_{g\bot}} = - \frac{GMr}{(\sqrt{D^2+r^2}^2+r_{\rm{plum}}^2)^{3/2}}\hat{\vec{r}},
\end{equation}
where $D =0.35$, about one half of the distance between our Galaxy and M31. The parallel force is

\begin{equation}
\vec{F_{g\|}}= - \left[ \frac{GM(r+D)}{((r+D)^2+r_{\rm{plum}}^2)^{3/2}} 
+ \frac{GM(r-D)}{((r-D)^2+r_{\rm{plum}}^2)^{3/2}}\right] \hat{\vec{r}}.
\end{equation}

{}

\end{document}